\documentclass[12pt,aps,prb,showpacs,preprint]{revtex4}   

\usepackage{amsmath}    
\usepackage{graphicx}   

\draft
\begin{document}
\newcommand{\be}{{\begin{equation}}}
\newcommand{\ee}{{\end{equation}}}
\newcommand{\la}{{\langle}}
\newcommand{\ra}{{\rangle}}
\newcommand{\eps}{{\epsilon}}
\newcommand{\lam}{{\lambda}}
\newcommand{\del}{{\delta}}
\newcommand{\Del}{{\Delta}}
\newcommand{\eq}{{\equiv}}
\newcommand{\nn}{{\nonumber}}
\newcommand{\Ga}{{\Gamma}}
\title{Understanding the Fano Resonance : through Toy Models}

\author{Swarnali Bandopadhyay$^\dagger$, Binayak Dutta-Roy$^\dagger$ 
and H.S.Mani$^*$}
 \affiliation{$^\dagger$S.N.Bose National Centre for Basic Sciences, J.D.Block, Sector-III, Salt Lake, Kolkata-98\\
\\
$^*$Institute of Mathematical Sciences, C.I.T. Campus, Taramani, Chennai 600113}

\begin{abstract}
The Fano Resonance, involving the mixing between a quasi-bound `discrete' state
of an inelastic channel lying in the continuum of scattering states belonging
to the elastic channel, has several subtle features. The underlying ideas have
recently attracted attention in connection with interference effects in quantum wires and mesoscopic transport phenomena. Simple toy models are provided in the present study to illustrate the basics of the Fano resonance in a simple and
tractable setting.
\end{abstract}
\maketitle

\section{Introduction}
Electronic states of atoms and molecules are characterized in the lowest
level of approximation (the Mean Field or Hartree-Fock description) in 
terms of electron occupancy of single particle states corresponding to
motion in an average potential. Thus, for example, the ground state of the
Helium atom is described as $(1s)^2$ $^1S_0$ viz two electrons in the spin singlet state in the lowest single particle orbital (in conformity with the Pauli exclusion principle). The first excited state is that of the excited configuration 
expressed in the standard spectroscopic notation as $(1s 2s)$ $^3S_1$ and so on.
Various excited states of the Helium atom may be excited from the ground state 
through the scattering of electrons or photons from the Helium atom and would 
show up as peaks (or Breit-Wigner resonances) in the scattering cross-section,
as also via the excitation spectrum mapping the inelastic processes. Of course
 our description through the electronic configuration is only approximate and
the true state is better described theoretically by considering the effect of
perturbations caused by the interactions beyond the mean field resulting in the
mixing of configurations. A qualitatively different situation arises, however, 
when we consider the excitation of a sufficiently high lying (energy wise) 
configuration, say for instance the $(2s2p)$ $^1P_1$ state of the Helium atom
which possesses an energy above the lowest ionization threshold of Helium. Such
a state if excited would be auto-ionizing viz. would decay into $He^+ + e^-$.
 Indeed, these manifest themselves as highly asymmetric peaks in the 
excitation spectrum. A qualitative understanding of this phenomenon was 
provided by Rice \cite{rice} and by Fano \cite{fano}. Quantitative treatment 
of the process, rendered somewhat subtle by the fact that it involves the
mixing between a discrete state and a continuum (with an attendant degeneracy
in energy) had to await almost a quarter of a century \cite{fano61}. 
 
There has been renewed interest in the Fano resonance in recent years in 
connection with interference effects in the quantum dot \cite{QD} and in the
Aharonov-Bohm ring with a quantum dot embedded in one arm\cite{AB-QD}. 
 Accordingly, it is desirable to 
expose the essence of the mechanism underlying the Fano resonance through 
exactly soluble models avoiding complications existing in realistic systems 
inessential for a basic understanding of the highly asymmetric excitation 
spectrum for a quasi-bound state lying in the continuum. In order to obtain 
a simple realisation of such a situation consider the elastic scattering of 
a particle by a target system describable in terms of a potential $V_g$. 
Suppose that the target has an excited state of energy $\Delta$ above its
ground state. When the projectile energy $E$ is above the threshold $\Delta$
then, apart from the elastic scattering the inelastic process is also possible.
One may describe the situation in terms of two channels, the elastic and the 
inelastic. The latter channel corresponds to the scattering of the projectile 
of energy $E > \Delta$ from the excited state of the system through a potential $V_e$, while the two channels are coupled to 
each other via a potential $V_c$. With $|\psi\ra$ and $|\phi\ra$ denoting the 
states in the elastic and inelastic channels respectively, the system is governed by the coupled Schr\"odinger equations    

\begin{subequations}
\begin{eqnarray}
(T + V_g) |\psi\ra + V_c |\phi\ra = E |\psi\ra, \label{eq:sch1}\\
(T + V_e) |\phi\ra + V_c |\psi\ra = (E-\Delta) |\phi\ra, \label{eq:sch2} 
\end{eqnarray}
\end{subequations}
where $T$ is the kinetic energy operator for the projectile. Now if the
projectile and excited target system interacting via the potential $V_e$
has a bound state whose energy will naturally be below $\Delta$ then we do
have a situation where the would-be bound state of the excited channel lies 
in the
continuum of the elastic channel, the desideratum for a Fano resonance. 

\begin{figure}[h]
\begin{center}
\resizebox{16cm}{16cm}{\includegraphics{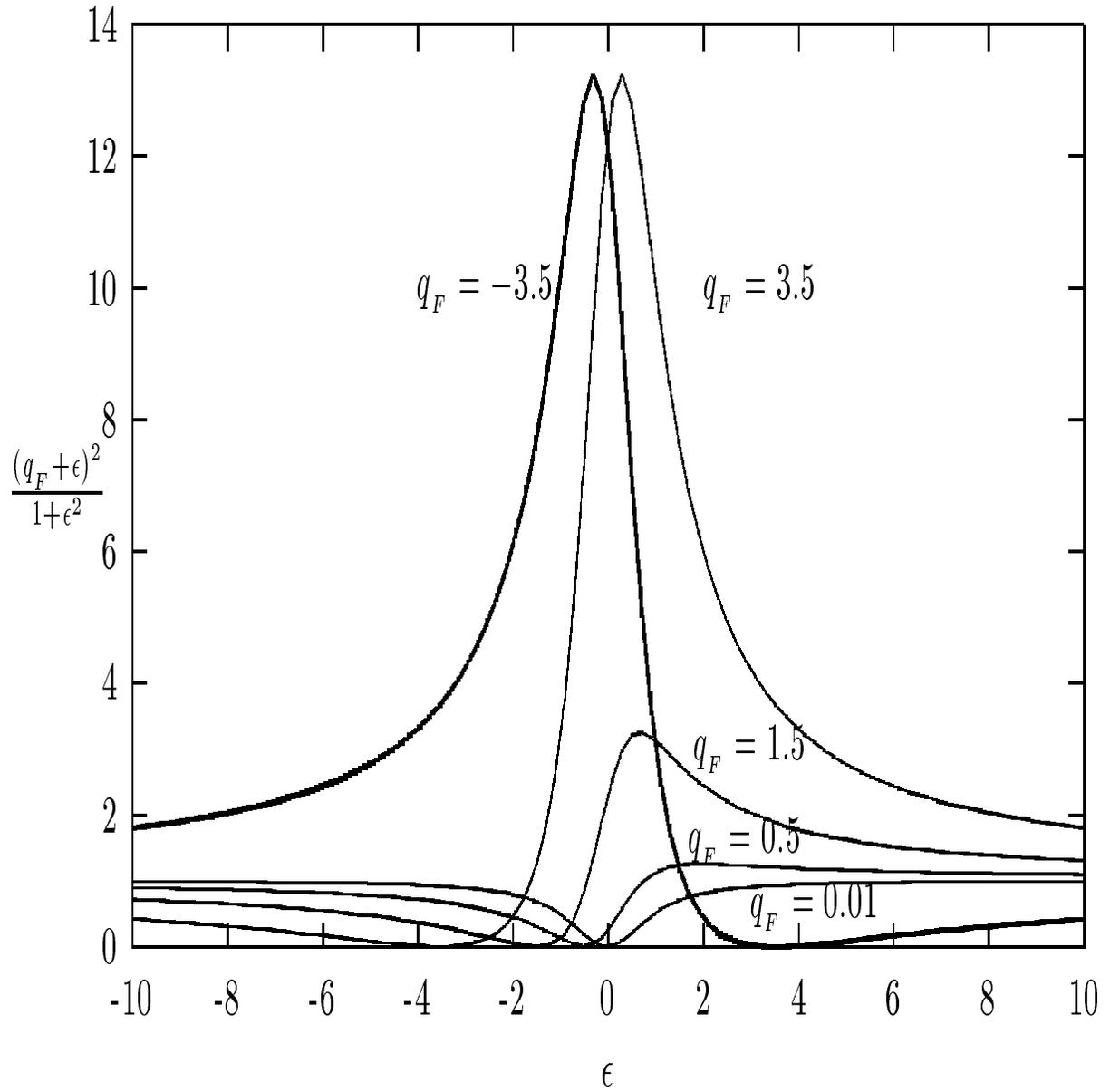}}
\caption{\label{fig:fano1}Fano line-shape for different values of Fano parameter $q$}
\end{center}
\end{figure}

We shall consider three illustrations by choosing the underlying potentials
and show that the probability for the excitation of the quasi-bound 
auto-ionizing state (so to say) exhibits a resonance at some energy 
$E_p < \Delta$ with a finite width $(\Gamma)$ corresponding to a complex pole
of the corresponding amplitude of the form $( E - E_p +i \Gamma/2 )^{-1}$
similar to what occurs in Breit-Wigner resonance. However, that apart the 
amplitude also possesses a concomitant zero at some real energy $E = E_0$
near the resonance energy $E_p$. It is the juxtaposition of a complex pole 
and a real zero of the excitation probability in the energy plane which gives 
rise to the observed 
highly asymmetric peak as the energy passes through that corresponding to
the auto-ionizing level. The generic pole-zero structure of the amplitude 
can be encapsulated in the form 
$\frac{E - E_0}{E - E_p + i \Gamma/2}= \frac{\eps +  q_{_F}}{\eps + i}$
with $ \eps $ measuring the separation of the energy from the resonance 
position in units of 
$\Gamma/2$ viz. $\eps = \frac{(E - E_p)}{\Gamma/2}$ (the reduced energy 
variable) and the asymmetry described by the Fano parameter 
$q_{_F} = \frac{(E_p - E_0)}{\Gamma/2}$.
Whence the excitation line shapes in the neighbourhood of the Fano resonance 
is given by
\begin{equation}
f(\eps) = \frac{(q_{_F} +\eps )^2}{1 +\eps ^2} 
\end{equation}
wherein to emphasize the main features the other weakly varying factors have 
been appropriately factored out. These line shapes \cite{fano61} for 
different values of 
the asymmetry parameter $ q_{_F} $ are shown in Fig.~\ref{fig:fano1}. The 
asymmetry parameter $ q_{_F} $ may be positive or negative depending on the 
relative disposition of the pole and the zero viz. the sign of $(E_p - E_0)$. 
The curves for negative values of $ q_{_F} $ may be obtained by changing 
$\eps \rightarrow -\eps $.

\section{Delta function model for the Fano-resonance}
We consider a soluble realisation of the two channel problem with a particle 
of mass $m$ moving in one dimension where we take 
$V_g(x) = 0$ while $V_c(x) = -\lam _c \del(x)$ and $V_e(x) = -\lam  \del(x)$
with $\lam > 0$. Taking $V_g(x) = 0$ is tantamount to saying that scattering 
in the elastic channel is only due to its coupling to the inelastic channel; 
this extreme simplification is to set forth in bold relief the Fano mechanism
as in such a situation the bound state in the continuum plays a central role
since but for the coupling between channels there is no effect (scattering)
 in the elastic channel.
Defining ${{2m\lam }\over {\hbar ^2}}\,\eq\, {\tilde \lam }$, ${{2m\lam _c}
\over {\hbar ^2}}\,\eq\, {\tilde \lam _c}$, ${{2mE }\over {\hbar ^2}}
\,\eq\, k^2$ and ${{2m(E-\Del)}\over {\hbar ^2}}\,\eq\, q^2$ the system 
equations [vide Eq.~(\ref{eq:sch1}) \& Eq.~(\ref{eq:sch2})] become
\begin{subequations}
\begin{eqnarray}
\frac{d^2}{dx^2} \psi(x) + {\tilde \lam _c } \del (x) \phi (x) = 
-k^2 \psi(x), \label{eq:sch1n}\\
\frac{d^2}{dx^2} \phi(x) + {\tilde {\lam }} \del (x) \phi (x) + 
{\tilde \lam _c }\del (x) \psi(x)= -q^2\phi(x), \label{eq:sch2n} 
\end{eqnarray}
\end{subequations}
We begin by discussing the region above the inelastic threshold ($\Del$) viz
we have $E>\Del$. Since the Fano paper \cite{fano61} discusses the entire 
problem using the partial wave phase shifts we also prefer to cast the one
dimensional transmission and reflection solutions in terms of eigen-channels
\cite{bayman} (analogous to angular momentum partial waves for central 
potentials see Appendix A). Noting that our equations of motion 
Eq.~(\ref{eq:sch1n}) 
\& Eq.~(\ref{eq:sch2n}) are symmetric under the operation $x \rightarrow -x$ 
we deduce that our solutions $\psi$ \& $\phi$ may be classified into sets that 
are even or odd under the transformation. With an attractive delta function
 potential in one dimension there is always one \& only one bound state which 
is in fact symmetric under parity. Also the odd ( or anti-symmetric ) function 
vanishes at $ x = 0$ \& hence the delta function potentials are ineffective. 
Therefore
the only nontrivial solutions of our problem are the ones that are even under
$x \rightarrow -x$. We observe that for $\lam _c = 0$ ( when two channels are
uncoupled ) the elastic channel corresponds to a free particle and has only
plane wave solutions, while in the inelastic channel we have continuum 
solutions ( taken to be box-normalised ) described by
\begin{subequations}
\begin{eqnarray}
\phi _0 (x) &=& \frac{1}{\sqrt{L}} [e^{-iqx} + a_0 e^{iqx} ]
\label{eq:2prop}\\
\hspace{-5.5cm}\text{\& \, a bound state solution}\hspace{1cm} 
\phi _0 (x) &=& \sqrt{\beta _0} e^{-\beta _0 |x|} \label{eq:2evans} 
\end{eqnarray}
\end{subequations}
\begin{equation}
\text{with} \,\, a_0 = \frac{q + i{\tilde \lam}/2}{q - i{\tilde \lam}/2}
\,\,\, \text{and} \, \, \beta _0 = \frac{{\tilde \lam}}{2}
\label{eq:2evansb0}
\end{equation}
We only consider the nontrivial solutions even under $x \rightarrow -x$.
The bound state occurs at $E = \Del - \frac{\hbar^2}{2 m}\,
\frac{{\tilde \lam}^2}{4} $. If 
$ \Del > \frac{\hbar^2}{2 m}\,\frac{{\tilde \lam}^2}{4} $, the bound state 
in the inelastic channel 
(lying of course below its threshold ) is in the region wherein the continuum
corresponding to elastic channel occurs. Also the bound state corresponds
to a pole at $ q = i {\tilde \lam}/2 $ of $a_0$.

Now let us consider the effect of coupling between the two channels. With
particles incident in the elastic channel we must solve Eq.~(\ref{eq:sch1n})
 \& Eq.~(\ref{eq:sch2n}) subject to the behaviour
\begin{subequations}
\begin{eqnarray}
\psi(x) = \frac{1}{\sqrt{4\pi}}[e^{-ik|x|} + A e^{ik|x|}] 
\label{eq:soln1}\\
\text{and} \hspace{2cm} \phi(x) = \frac{a}{\sqrt{2\pi}}e^{iq|x|}
\label{eq:soln2}
\end{eqnarray}
\end{subequations}
since in the inelastic channel we have only outgoing waves. We obtain
\begin{subequations}
\begin{eqnarray}
A = \frac{q - i\frac{{\tilde \lam}}{2} - \frac{{\tilde \lam _c}^2}{4k}}
{q - i\frac{{\tilde \lam}}{2} + \frac{{\tilde \lam _c}^2}{4k}} 
\label{wavampl-elas}\\
a = \frac{i {\tilde \lam _c}}{\sqrt{2}(q-i\frac{{\tilde \lam}}{2}
+ \frac{{\tilde \lam _c}^2}{4k})}\label{wavampl-inelas}
\end{eqnarray}
\end{subequations}
and $|A|^2 + 2|a|^2 \frac{q}{k} = 1$ , 
which follows directly from Eq.~(\ref{wavampl-elas})
 \& Eq.~(\ref{wavampl-inelas}) and is a consequence of unitarity with 
respect to the 
two channels elastic and inelastic. The factor of $\frac{q}{k}$ arising
from the ratio of the currents in the inelastic and elastic channels and
the factor of $2$ representing particles in the inelastic channel emerging
left-wards and right-wards. We may also observe that the pole of $A$ has 
shifted as compared to that of $a_0$
as an effect of the coupling embodied by $\lam _c$ and to find its location
we have to solve a quartic equation in $k$. We may also note that as 
$\lam _c \to 0 ,\,\, A \to 1 $ as it should since for simplicity we
have taken $ V_g = 0 $. Rather than resorting to numerical
results let us solve for the pole position ( $k_p$ ) for small $\lam _c$.
We find readily (see appendix B ) that
\begin{equation}\label{eq:pole-del-k}
k_p \cong \sqrt{{\tilde \Del} - \frac{{\tilde \lam}^2}{4}}\,
+ {\tilde \lam _c}^4 \, \frac{(2\,{\tilde \Del}\,+\,{\tilde \lam}^2)}
{64({\tilde \Del} - \frac{{\tilde \lam}^2}{4})^{5/2}}\, 
- i\,{\tilde \lam _c}^2 \,\frac{{\tilde \lam}}
{8({\tilde \Del} - \frac{{\tilde \lam}^2}{4})}
\end{equation} 
Thus for small $\lam _c$ the pole moves off from the real axis in the 
$k$-plane and sits in the fourth quadrant close to the real axis. Therefore,
when we map onto the energy $k^2$-plane this pole vis a vis the unitarity
cut ( with branch point at $ E = 0 $) lies in the fourth quadrant of the
second Riemann sheet, as befits a resonance and accordingly
\begin{equation}\label{eq:pole-del-E}
E_p - i \frac{\Ga}{2} \cong 
\Del - \frac{\hbar^2}{2 m}\,\left[\frac{{\tilde \lam}^2}{4}
- {\tilde \lam _c}^4\,\frac{({\tilde \Del} + \frac{{\tilde \lam}^2}{4})}
{16({\tilde \Del} - \frac{{\tilde \lam}^2}{4})^2}\right]
- i \,\frac{\hbar^2}{2 m}\, {\tilde \lam _c}^2 \,\frac{{\tilde \lam}}
{4({\tilde \Del} - \frac{{\tilde \lam}^2}{4})^{1/2}} 
\end{equation}   
For a physical understanding of the width $(\Ga)$ of the resonance we recognize
that this `bound state in a continuum' or the `auto-ionizing' state having an 
unperturbed energy $E = \Del - \frac{\hbar^2}{2 m}\,\frac{{\tilde \lam}^2}{4}$  
decays into the continuum states of the elastic channel caused by the 
interaction $H_{int} = - \lam _c \, \del(x)$. To obtain the rate of the decay
for small ${\tilde \lam _c}$ it would suffice to use time-dependent 
perturbation theory or the Fermi Golden Rule viz. 
$Rate = \frac{2\pi}{\hbar} \bigl | \la f |H_{int}| i \ra \bigr |^2 
\frac{d\rho}{dE}$ where initial state is the unperturbed bound state
described by the wave-function $\la x | i \ra = \sqrt{\frac{{\tilde \lam }}{2}} 
e^{-\frac{{\tilde \lam }}{2} |x|}$ and the final unperturbed state
is $\la x | f \ra = \frac{1}{\sqrt{L }} e^{ik|x|}$ (taking box normalisation
 say ) and the number of states of a free particle between momenta $p$ and 
$p + dp$ is $d\rho = \frac{L dp}{h}$, which gives 
$\frac{d\rho}{dE} = \frac{L}{h} \frac{d (\hbar k)}{dE} = \frac{L}{2\pi} 
\bigl (\frac{d}{dE} \sqrt{\frac{2 m E }{\hbar^2}}\bigr )_{\! E = 
\Del - \frac{\hbar^2}{2m} \, \frac{{\tilde \lam}^2}{4}} 
= \frac{L}{4\pi} \, \sqrt{\frac{2 m}{\hbar^2}}\,
\frac{1}{\sqrt{\Del - \frac{\hbar^2}{2m} \,\frac{{\tilde \lam}^2}{4}}}$ .
 Accordingly, the decay rate for the `auto-ionizing' state to go into the
continuum is given by $ \frac{\Ga}{\hbar} = \frac{2\pi}{\hbar} \, \, 
\bigl| \int_{-\infty}^{\infty} \frac{1}{\sqrt{L}} \, e^{-ik|x|} \, 
(-\lam _c) \del(x) \sqrt{\frac{{\tilde \lam}}{2}} 
\, e^{-\frac{{\tilde \lam }}{2}|x|} \, 
dx \, \bigr|^2 \, \frac{L}{4\pi} \, \sqrt{\frac{2 m}{\hbar^2}}\,\frac{1}
{\sqrt{\Del - \frac{\hbar^2}{2m} \,\frac{{\tilde \lam}^2}{4}}}
 = \frac{\hbar}{2 m }\,\frac{{\tilde \lam _c}^2 {\tilde \lam }}
{4 \,\sqrt{{\tilde \Del} - \frac{{\tilde \lam}^2}{4}}} $ 
in agreement with our earlier result 
describing the excursion of the pole into the complex plane.
Further one may use perturbation theory to verify that the shift in the real 
part of energy to order ${\tilde \lam _c}^2$ is zero.

In order to understand the origin of the Fano zero we must examine the situation
 at energies below the onset of the inelastic channel $ E < \Del$. In this region the system equations become,
\begin{subequations}
\begin{eqnarray}
\frac{d^2}{dx^2} \psi(x) + {\tilde \lam _c } \del (x) \phi (x) = 
-k^2 \psi(x), \label{eq:sch1nn}\\
\frac{d^2}{dx^2} \phi(x) + {\tilde \lam } \del (x) \phi (x) + 
{\tilde \lam _c }\del (x) \psi(x)= \beta^2\phi(x), \label{eq:sch2nn}
\end{eqnarray}
\end{subequations}
when $\beta^2=\frac{2m}{\hbar^2}(\Del-E) > 0$. It needs to be emphasized 
that $\beta$ here is not an eigenvalue to be determined but it is fixed 
given the energy $ E $ of the incident beam. The open channel solutions 
$ \psi(x) $ correspond to continuum states while the closed or evanescent 
channel function $ \phi(x) $ is exponentially damped. However, since we are 
concerned with the mixing between $ \psi(x) $ \& $ \phi(x) $ pertaining
to a bound state in the continuum we must adopt a scheme for the normalisation
of the two functions treated at par. This is indeed what is proposed by Bayman
\& Mehoke \cite{bayman} where the open channels are normalized to unit flux 
and the closed channels with an analogous factor (which would result from 
analytic continuation). Thus with $\xi_k$ \& $\sqrt{1-|\xi_k|^2}$ being the
probability amplitudes (or mixing coefficients) to be determined ,
\begin{subequations}
\begin{eqnarray}
\psi_k(x) = \xi_k \, \frac{1}{\sqrt{4 \pi}} \,\frac{1}{\sqrt{k}}\, 
\bigl(e^{-ik|x|}+Ae^{ik|x|}\bigr) \,, 
\label{eq:soln1nn}\\
\phi(x) = \sqrt{1-|\xi_k|^2} \, \frac{1}{\sqrt{i\,\beta}} \, 
\frac{1}{\sqrt{(1 - e^{-\beta L})}}\,
e^{-\beta |x| } \label{eq:soln2nn}
\end{eqnarray}
\end{subequations}
which when substituted in the system equations (\ref{eq:sch1nn} \& 
\ref{eq:sch2nn}) yield
\begin{subequations}
\begin{eqnarray}
A &=& \frac{{\tilde \lam } - 2\beta - i\frac{{\tilde \lam _c }^2}{2k}}
{{\tilde \lam } - 2\beta + i\frac{{\tilde \lam _c }^2}{2k}} 
\label{eq:ealscoef}\\
\frac{ \xi_k }{ \sqrt{ 1 -|\xi_k|^2 }} &=& \frac{ \sqrt{ \pi k }}
{\sqrt{ \beta}} \,\frac{1}{{\tilde \lam _c }} \,e^{i\,\frac{\pi}{4}}\,
\left( 2\beta - {\tilde \lam } + \frac{{\tilde \lam _c }^2}{ 2ik } \right)
\label{eq:evcoef}
\end{eqnarray}
\end{subequations}
It may be observed that $|A| = 1$ in view of unitarity since the inelastic 
channel is closed (or evanescent).
Note that in the limit $\lam _c \to 0$ we have $ |\xi_k| \to 1 $ which means 
that the bound state decouples and probability to be in the elastic 
channel is unity. When we try to access the bound state embedded in the 
continuum through elastic scattering (or other processes ) the transition 
probability will involve the overlap integral. The probability for exciting 
the `auto-ionizing' state 
described via scattering is proportional to the square modulus of the
overlap integral between $ \phi $ \& $ \psi_k $ which is
\begin{equation}
\left| \,\int _{-\infty} ^\infty \phi^{\! *}(x) \, \psi_k(x) \, dx \right|^2 = 
2 \, \frac{ \left( 1 - |\xi|^2 \right)^2 }{ \beta ^2 + k^2 } \,
\left[ \frac{2\beta - {\tilde \lam }}{{\tilde \lam _c }} - 
 \frac{{\tilde \lam _c }}{2 \beta } \right]^2
\end{equation}
which is seen to vanish when
\begin{equation}
\beta ({\tilde \lam } - 2\beta ) + \frac{{\tilde \lam _c }^2}{2} = 0
\end{equation}
that is when $\beta = \frac{{\tilde \lam } + \sqrt{{\tilde \lam }^2 + 
4 {\tilde \lam _c }^2}}{4}$ . 
This is the celebrated Fano zero with the 
experimental signature that the line shape for excitation of the auto-ionizing
state vanishes at some energy 
$ E_0 $. Therefore, it may also be observed that the transmission amplitude,
$ A + 1 $ (see Appendix A) has a zero at ${\tilde \lam} = 2 \, \beta$.
To form a 
picture of the composite situation that has emerged it is worthwhile to look 
at this again 
for small ${\tilde \lam _c }$ whence we find that the Fano-zero is located 
at $ E_0 \simeq \Del - \frac{\hbar^2}{2 m }\,\left[\frac{{\tilde \lam }^2}{4}  
+ {\tilde \lam _c }^4 \, \frac{({\tilde \Del } + \frac{{\tilde \lam }^2}{4})}
{16 ({\tilde \Del} - \frac{{\tilde \lam }^2}{4} )^2 }\right]$
 as compared to the resonance
energy $E_p = \Del - \frac{\hbar^2}{2 m}\,\left[\frac{{\tilde \lam }^2}{4} 
- {\tilde \lam _c }^4 \, \frac{({\tilde \Del } + \frac{{\tilde \lam }^2}{4})}
{16 ({\tilde \Del} - \frac{{\tilde \lam }^2}{4} )^2 }\right]$ 
and the width $\Ga = \frac{\hbar^2}{2 m }\,
\frac{{\tilde \lam }{\tilde \lam _c }^2}
{2 \sqrt{{\tilde \Del} -\frac{{\tilde \lam }^2}{4}}}$  with the Fano parameter
$q_{_F} > 0 $. 
It is evident from the above discussion that the essence of the Fano zero
lies in the mixing between the continuum states of the elastic channel and 
quasi-bound `discrete' state of the inelastic channel. This is well expressed
by Ugo Fano \cite{fano68} : ``Repulsion ( of levels ) is familiar in discrete
spectra, when a level of one configuration happens to lie in the midst of a 
series of levels of other configurations; configuration interaction causes 
the levels of the second series to be shifted away from the perturbing level.
 In our situation, where a discrete level of one series lies in the continuum
of another channel, the levels of this continuum are also repelled, but in
the sense that their oscillator strength (level densities) is thinned out in 
the proximity of the perturbing level" (the addition under quotes given in
parentheses are ours and are provided in the interest of increased clarity) .
To concretise the remark in the present context we observe the effect of the 
coupling on the density of states (DOS) of the elastic channel. 
The changed DOS due to the ``auto-ionizing state" is thus given by the 
multiplicative factor
$$
\left| \xi_k \right|^2 = 1 - \frac{\frac{\beta {\tilde \lam_c }^2}{\pi k}}
{( 2 \beta - {\tilde \lam} )^2 + \frac{{\tilde \lam_c }^2}{4 k^2}
+ \beta \frac{{\tilde \lam_c }^2}{\pi k}}  \le  1
$$
thus reducing the density of levels via ``level repulsion".
\section{Spherical Dirac delta shell model for Fano resonance}
\label{sec2}
To provide a three dimensional version of the problem at hand we consider a
spherical shell type coupling potential $[-\lam_c \del (r-a)]$ between the
two channels and the same form $[-\lam \del (r-a)]$ for the potential operative
in the inelastic channel. We confine ourselves to low energy scattering so only
the $l=0$ partial wave need be taken into account. Defining 
${{2m\lam }\over {\hbar ^2}}\,\eq\, {\tilde \lam }$, ${{2m\lam _c}
\over {\hbar ^2}}\,\eq\, {\tilde \lam _c}$, ${{2mE }\over {\hbar ^2}}
\,\eq\, k^2$ and ${{2m(E-\Del)}\over {\hbar ^2}}\,\eq\, q^2$
the system equations [vide Eq.~(\ref{eq:sch1}) \& Eq.~(\ref{eq:sch2})] become 
\begin{subequations}
\begin{eqnarray}
\frac{1}{r^2} \, \frac{d}{dr}[r^2 \frac{d\psi(r)}{dr}] + 
{\tilde \lam _c} \del(r-a) \phi(r) = -k^2 \psi(r)\label{sec2sch1}\\ 
\frac{1}{r^2} \, \frac{d}{dr}[r^2 \frac{d\phi(r)}{dr}] + 
{\tilde \lam} \del(r-a)\phi(r) +  
{\tilde \lam _c} \del(r-a)\psi(r) 
=-q^2 \phi(r) \label{sec2sch2}
\end{eqnarray}
\end{subequations}
Using $\psi(r)= \frac{v(r)}{r}$ and $\phi(r)= \frac{u(r)}{r}$ in 
Eq.~(\ref{sec2sch1}) \& Eq.~(\ref{sec2sch2}), the system equations 
are transformed to
\begin{subequations}
\begin{eqnarray}
\frac{d^2v(r)}{dr^2} + {\tilde \lam _c} \del(r-a) u(r) 
= -k^2 v(r)\label{sec2elas}\\ 
\frac{d^2u(r)}{dr^2} + {\tilde \lam} \del(r-a)u(r) +  
{\tilde \lam _c} \del(r-a)v(r) 
=-q^2 u(r) \label{sec2inelas}
\end{eqnarray}
\end{subequations}
We observed that when two channels are uncoupled (i.e. $\lam _c = 0$),
similar to previous case, the elastic channel corresponds to free particle 
while inelastic channel can be described by continuum solutions and a bound 
state. In presence of coupling between two channels we must solve 
Eq.~(\ref{sec2elas}) \&
 Eq.~(\ref{sec2inelas}) subject to behaviour
\begin{subequations}
\begin{eqnarray}
v(r) \>=\> \left\{ \begin{array}{lcc} A(e^{ikr} - e^{-ikr}), \, \, 0<r<a \\
C e^{ikr} + D e^{-ikr}, \, \, a<r \end{array}\right.
\end{eqnarray}
\begin{eqnarray}
u(r) \>=\> \left\{ \begin{array}{lcc} X(e^{iqr} - e^{-iqr}), \, \, 0<r<a \\
Y e^{iqr}, \, \, a<r \end{array}\right. 
\end{eqnarray}
\end{subequations}
describing particle incident in the elastic channel. Note that $v(r=0) = 0$ and
$u(r=0) = 0$ as required for well behaved wave functions.
Using proper boundary conditions for this system 
(continuity of the wave functions and jump condition for the derivatives)
we obtain
\begin{subequations}
\begin{eqnarray}
\frac{C}{D} = \frac{\frac{q}{Sin(qa)} - {\tilde \lam} e^{iqa} 
-\frac{{\tilde \lam _c}^2}{k} Sin(ka) e^{i(q-k)a}}
{-\frac{q}{Sin(qa)}+{\tilde \lam} e^{iqa}+\frac{{\tilde \lam _c}^2}{k} Sin(ka) 
e^{i(q+k)a}}\\
\frac{Y}{D} = \frac{2i{\tilde \lam _c}Sin(ka)}
{-\frac{q}{Sin(qa)}+{\tilde \lam} e^{iqa}+\frac{{\tilde \lam _c}^2}{k} Sin(ka) 
e^{i(q+k)a}}
\end{eqnarray}
\end{subequations}
We find that $ |\frac{C}{D}|^2 + |\frac{Y}{D}|^2 \, \frac{q}{k} = 1$ which is 
a consequence of unitarity with respect to the two channels elastic and 
inelastic. As was the case for the 1D model we can see that the poles of 
${C}\over{D}$ has shifted in presence of the coupling between two channels.

When the inelastic channel is closed (evanescent) i.e. $ E < \Del $ 
or $ q $ is purely imaginary (say $ q = i \beta $ ) then
\begin{subequations}
\begin{eqnarray}
\frac{C}{D} = \frac{\frac{\beta}{Sinh(\beta a)} - {\tilde \lam} e^{-\beta a} 
-\frac{{\tilde \lam _c}^2}{k} Sin(ka) e^{-(\beta + ik) a}}
{-\frac{\beta }{Sinh(\beta a)}+{\tilde \lam} e^{-\beta a}+
\frac{{\tilde \lam _c}^2}{k} Sin(ka) e^{-(\beta - ik)a}}\label{soln-elas}\\
\frac{Y}{D} = \frac{2i{\tilde \lam _c}Sin(ka)}
{-\frac{\beta }{Sinh(\beta a)}+{\tilde \lam} e^{-\beta a}+
\frac{{\tilde \lam _c}^2}{k} Sin(ka) e^{-(\beta - ik)a}}\label{soln-inelas}
\end{eqnarray}
\end{subequations}
and the bound-state of the inelastic channel can be degenerate with the 
continuum of the elastic channel. From Eq.~(\ref{soln-elas}) and 
Eq.~(\ref{soln-inelas}), when two channels are uncoupled 
(i.e. $ \lam _c = 0 $), 
 we observe that $ C = -D $ and $ Y = 0 $ which imply that everything is 
transmitted out through the elastic channel. The square modulus of the overlap 
integral between the closed inelastic channel $ \phi $ and open elastic channel
$ \psi $ is 
\begin{equation}
\left| 4\pi \int_0^{\infty} u^*(r) v(r) dr \right|^2 = 
\frac{16\pi |XA|^2}{(\beta ^2 + k^2)^2} \left[ 2\beta e^{\beta a} - 
\frac{4 {\tilde \lam _c}^2 e^{-\beta a}Sinh^2(\beta a)}
{2\beta + {\tilde \lam} (e^{-2 \beta a} -1)}\right]^2 Sin^2(ka)
\end{equation}
which is seen to vanish when
\begin{equation}
\beta \left[ ({\tilde \lam} - 2 \beta ) - {\tilde \lam} e^{-2 \beta a }\right]
+ \frac{{\tilde \lam _c}^2 }{2} \left[ 1 - e^{-2 \beta a } \right]^2 = 0
\end{equation}
Since this is a transcendental equation one should take recourse to numerical 
calculations, but we choose for simplicity to consider the situation where 
$\lam _c$ is small and $\beta a$ is small; hence ${\tilde \lam} a $ is close 
to unity.
In such a circumstance we find that the overlap integral vanishes when
\begin{equation}
\beta = 1 - \frac{1}{{\tilde \lam} a} + 
\frac{{\tilde \lam _c}^2 a}{{\tilde \lam}}
\end{equation} 
or at an energy
\begin{equation}
E_0 \simeq \Del -\frac{\hbar^2}{2m}\left[  
\frac{1}{a^2} (1-\frac{1}{{\tilde \lam} a})^2 +
\frac{2}{{\tilde \lam} a}{\tilde \lam _c}^2 (1-\frac{1}{{\tilde \lam} a})
+ \frac{{\tilde \lam _c}^4}{{\tilde \lam }^2}\right]
\end{equation}
which is the ``Fano zero".
\section{ Non-local Separable Model for Fano-resonance} 
To provide another soluble example, this time again in three dimensions, to 
illustrate 
the notion of the Fano resonance, we introduce the idea of a non-local 
potential. To make this article self-contained for begining students
we reconsider the Schr\"odinger equation for a stationary state 
$\bigl( T + V \bigr) |\psi \ra = E |\psi \ra $ for a
particle of mass $ m $ moving in a potential $ V $ is in the coordinate 
representation $ -\frac{\hbar^2}{2m} \nabla^{\!2} \psi(\vec r) \, 
+\,\la \vec r | V| \psi \ra = E \psi(\vec r)$. Now $ \la \vec r | V| \psi \ra = 
\int \la \vec r | V| \vec r' \ra \, \la \vec r' | \psi \ra \, d^3\vec r'
 = \int \la \vec r | V| \vec r' \ra \, \psi(\vec r') d^3\vec r' $ 
where we have simply used the completeness of the states $| \vec r \ra $.
Usually we deal with local potentials where $ \la \vec r | V| \vec r' \ra 
= V(r) \del (\vec r - \vec r')$ and the potential term simply reduces 
to $ V(r) \psi (\vec r )$. However, in general, since a particle in Quantum 
Mechanics is not necessarily localized we may admit non-local potentials. 
Furthermore, if we take a non-local but separable and central potential 
$\la \vec r | V| \vec r' \ra = -\alpha \, \frac{e^{-\mu r}}{r} \,
\frac{e^{-\mu r'}}{r'} $ where we have also chosen a specific (Yukawian) form
of the radial dependence for illustration, we have several simplifications.
Firstly the potential is effective only in the $ l =0 $ state because only
that component survives the onslaught of the angular integration in
$ \int \la \vec r | V| \vec r' \ra \, \psi(\vec r') d^3\vec r' $. 
Secondly the solutions of the equation involve nothing more complicated than 
exponential functions. Thus choosing a model analogous to that described by the
system equations (~\ref{sec2elas}) \& (~\ref{sec2inelas}) except that now we 
have in place of delta functions the non-local separable potentials of 
the chosen form, we are invited to consider the equations
\begin{subequations}
\begin{eqnarray}
\frac{d^2v(r)}{dr^2} \, + {\tilde \alpha_c} \, e^{-\mu r} \, \int^{\infty}_0 
e^{-\mu r'} \, u(r') \, dr' = -k^2 v(r) \label{eq:nonloc1}\\
\frac{d^2u(r)}{dr^2} \, + {\tilde \alpha} \, e^{-\mu r} \, \int^{\infty}_0 
e^{-\mu r'} \, u(r') \, dr' + {\tilde \alpha_c} \, e^{-\mu r} \, \int^{\infty}_0
e^{-\mu r'} \, v(r') \, dr' = -q^2 u(r) \label{eq:nonloc2}
\end{eqnarray}
\end{subequations}
where we have written $\psi(r) = \frac{v(r)}{r}$ and 
$\phi(r) = \frac{u(r)}{r}$. The other symbols have meanings completely analogous to what we had earlier viz. $k^2 = \frac{2m}{\hbar^2} E$, 
$q^2 = \frac{2m}{\hbar^2} (E-\Del)$, 
${\tilde \alpha} = 4 \pi \frac{2m \alpha}{\hbar^2}$, 
${\tilde \alpha_c} = 4 \pi \frac{2m \alpha_c}{\hbar^2}$.
We have written the system equations for energies $( E > \Del )$ viz.
above the threshold where both channels are open. First putting $\alpha_c = 0 $
we observe that the elastic channel corresponds to a free particle 
(so chosen as to avoid unnecessary complications we have, as earlier,
taken the elastic channel to be non-trivial only through its coupling
to the inelastic channel). On the other hand the inelastic channel has 
continuum solutions described by
\begin{equation}
u_0(r)  = e^{-iqr}\,- s_0\,e^{iqr}\,+ (s_0 - 1 )\,e^{-\mu r} 
\label{eq:nonloc2sol}
\end{equation}
corresponding asymptotically to incoming and outgoing 
spherical waves (the latter
suffering a phase-shift $ s_0 = e^{2 i \del_0}$ ) and the last 
term inserted to guarantee that $u_0( r 
= 0 ) =0, [$ as $\psi(r)= \frac{u(r)}{r}]$ and this is needed to guarantee 
the well-behavior of the wave function. Substituting Eq.(~\ref{eq:nonloc2sol}) 
in Eq.(~\ref{eq:nonloc2}) setting $\alpha_c=0$ we get
\begin{subequations}
\begin{eqnarray}
s_0 = \frac{\mu^2 + q^2 -\frac{{\tilde \alpha}}{2 \mu}\,
\frac{\mu - i q}{\mu + i q}}{\mu^2 + q^2 -\frac{{\tilde \alpha}}{2 \mu}\,
\frac{\mu + i q}{\mu - i q}}\hspace{6cm}\\
\text{As before the bound state correspond to the pole of the  } S_0 \nn 
\text{ and occurs at}\hspace{2.4cm}\nn\\
q=i\beta_0=i\bigl(\sqrt{\frac{{\tilde\alpha}}{2\mu}}-\mu\bigr)\hspace{6cm} 
\end{eqnarray}
\end{subequations}
provided ${\tilde \alpha } > 4 \mu^3 $ \& the potential is sufficiently strong
to support a bound state.

Now let us consider the effect of the coupling between the two channels.
 With particle incident in the elastic channel we must solve equations 
(\ref{eq:nonloc1}) \&  (\ref{eq:nonloc2}) subject to the behaviour
\begin{subequations}
\begin{eqnarray}
v(r) = e^{-ikr} - S e^{ikr} + ( S - 1 ) e^{- \mu r}\\
u(r) = t e^{iqr} - t e^{-\mu r}
\end{eqnarray}
\end{subequations}
with only outgoing asymptotic waves in the inelastic channel, and of course
$ v( r=0 ) = 0 $ and $ u( r=0 ) = 0 $. Substitutions into the equations 
(\ref{eq:nonloc1}) \&  (\ref{eq:nonloc2}) yield
\begin{subequations}
\begin{eqnarray}
S = \frac{( \mu - i q)^2 - \frac{{\tilde \alpha}}{2 \mu } - 
\bigl( \frac{{\tilde \alpha_c}}{2 \mu}\bigr)^2 / ( \mu + i k)^2}
{( \mu - i q)^2 - \frac{{\tilde \alpha}}{2 \mu } - 
\bigl( \frac{{\tilde \alpha_c}}{2 \mu}\bigr)^2 / ( \mu - i k)^2}\\
\nn\\
t = \frac{- \frac{2 i k }{\mu^2 + k^2 } {\tilde \alpha_c} \frac{ \mu - i q}
{\mu + i q}}{( \mu - i q)^2 - \frac{{\tilde \alpha}}{2 \mu } -
\bigl( \frac{{\tilde \alpha_c}}{2 \mu}\bigr)^2 / ( \mu - i k)^2}
\end{eqnarray}
\end{subequations} 
Again we can verify when $q$ is real that $|S|^2 + |t|^2 \frac{q}{k}$ and when 
$q=i\beta$, $\beta$ real $|S|=1$ which are consequences of inelastic and 
elastic unitarity respectively. However when $(\mu + \beta)^2 = 
\frac{{\tilde \alpha}}{2\mu} + \left(\frac{{\tilde \alpha_c}}{2\mu}\right)^2
\frac{\mu^2 - k^2}{(\mu^2 + k^2)^2}$ then $S = -1$.
As was the case for the Dirac delta model we can see that the poles of $S$ has shifted as compared to that of $S_0$ due to the coupling between the channels embodied by $\alpha_c$ and one similarly finds that the bound state of the inelastic channel becomes a resonance in the elastic channel corresponding to a pole located in the fourth quadrant of the second Riemann sheet in the complex energy plane.\\

In a similar fashion examining the situation at energies below the onset of the inelastic channel ($ E<\Delta$) we may study the mixing of the quasi-bound 
``auto-ionizing" state and the continuum and the probability for exciting this 
state via scattering to discover the energy at which the overlap integral 
vanishes ({\it viz.} the Fano zero)
\begin{eqnarray}
E_0 = \Delta - \frac{\hbar^2}{2m}\left[\mu^2 + \left(\frac{{\tilde\alpha}}{4\mu} + \frac{\sqrt{{\tilde\alpha}^2 + 4{\tilde\alpha_c}^2}}{4\mu}\right) 
-2\mu\left(\frac{{\tilde\alpha}}{4\mu}+ \frac{\sqrt{{\tilde\alpha}^2 + {\tilde\alpha_c}^2}}{4\mu} \right)^{\frac{1}{2}}\right]
\end{eqnarray}

\section{Discussion of results}

We have attempted to explain the origin of Fano resonance through three soluble 
models: with one-dimensional Dirac delta potential, with Dirac-delta shell 
potentials and with a  
non-local separable potential in three dimensions. To make the results more 
transparent we have taken the elastic channel to correspond to a free particle 
but for its coupling to the inelastic channel so that the bound state of the 
latter manifests itself through the typical Fano pole-zero structure. Of course
 one could have included a direct potential in the elastic channel as well but 
this would only complicate the algebra without adding anything to the 
understanding of the relevant mechanism. It has also been clarified how 
the amplitudes generate a pole in the fourth quadrant of the second Riemann 
sheet of the complex energy plane in the elastic unitary cut, and how the 
excitation functions for the ``auto-ionizing" quasi-bound state from the 
elastic channel develop a zero at some real energy near the resonance. The 
generic reason for the occurrence of this zero is explained by Fano as arising 
from the fact that the wave function in the elastic channel will be of the form 
$ e^{-ikr}-e^{2i\delta}e^{+ikr}$ which is $Sin(kr+\delta)$ modulo an over-all 
phase and irrelevant factors. To find the excitation probability of the 
quasi-bound state we have to determine the overlap of this with some smooth 
function $g(k,r)$ [ which varies slowly with respect to $k$]. Now 
$\int Sin(kr+\delta)\,g(k,r)dr=Cos\delta\int Sin kr\, 
g(k,r)dr +Sin\delta \int Cos kr\, g(k,r)dr$. As the energy crosses the Fano 
resonance $\delta$ changes from $0$ to $\pi$, the overlap integral must pass 
through zero. In order to make the one-dimensional model to also conform to 
this generic pattern we have adopted the formalism of eigen-channels to 
introduce therein the 
notion of a phase-shift. We have also demonstrated the modification in the 
density of states due to the ``bound state in the continuum".

It is perhaps necessary to add a word of caution. It may sometimes be very
difficult from the data to discriminate between a Fano-resonance \& the 
interference effects between a Breit-Wigner resonance \& a slowly varying 
background. Thus consider the latter situation described by an amplitude 
$ f = {\sl B}\, + \,\frac{\frac{\Ga}{2}}{E-E_r+i\frac{\Ga}{2}}$ and note that 
$ | f |^2 = {\sl B}^2 \,+\,\frac{\frac{\Ga^2}{4}\, +\, {\sl B} \Ga (E-E_r) }
{(E-E_r)^2 \,+\, \frac{\Ga^2}{4}} $. The interference term between the 
resonance \& the background could very well be parametrized by the Fano form 
with the asymmetry parameter $ q = \frac{E_p-E_0}{\frac{\Ga}{2}} = 
\frac{1}{{\sl B}}$. 
\section{acknowledgment}

We are grateful to Dr. Prosenjit Singha Deo who introduced us to the idea of the Fano resonance and encouraged us to search for simple illustrations to clarify the underlying mechanism.


\appendix
\section{Location of the transmission zero}
For scattering in three dimensions from spherically symmetric potentials 
the $ S $-matrix is diagonal in the orbital angular momentum partial wave 
channels \& hence unitarity forces each diagonal to be of the form 
$ e^{2i\del_l} $, where $ \del_l $ is the corresponding phase shift. For
a one dimensional problem with particles incident from the left 
$ \langle x | k \rangle = \frac{1}{\sqrt{2 \pi}}\,e^{ikx}$ we search for
solutions 
\begin{eqnarray} 
\psi (x) \,& \longrightarrow &
 e^{ikx} \,+ \,a_{_R}^{(L)}\, e^{-ikx} \, \, \, \, \, 
 \text{ as } x\to-\infty \nn\\
& \longrightarrow & a_{_T}^{(L)}\, e^{ikx} \, \, \, \, \, \,\text{ as } 
x\to+\infty \nn
\end{eqnarray}
where $ a_{_R}^{(L)} $ is the reflection amplitude and $ a_{_T}^{(L)} $ is the 
transmission amplitude for the left incident particles.
Similarly for particles incident from the right $\la x | -k \ra = \frac{1}
{\sqrt{2\pi}}\,e^{-ikx}$ we look for 
\begin{eqnarray}
\psi (x) \,& \longrightarrow &
 e^{-ikx} \,+ \,a_{_R}^{(R)}\, e^{ikx} \, \, \, \, \, \text{ as } x\to+\infty 
\nn\\
& \longrightarrow & a_{_T}^{(R)}\, e^{-ikx} \, \, \, \, \, \,
\text{ as } x\to-\infty \nn
\end{eqnarray}
where $ a_{_R}^{(R)} $ is the reflection amplitude and $ a_{_T}^{(R)} $ is the 
transmission amplitude for the right incident particles.
Thus the $ S $-matrix has the form
\begin{equation}
S\,=\,\left(\begin{array}{cc}
a_{_T}^{(L)} & a_{_R}^{(R)}\\
a_{_R}^{(L)} & a_{_T}^{(R)}
\end{array}\right)
\end{equation}
Now if $ V(x) $ enjoys the symmetry $ V(-x) = V(x) $ then clearly 
$ a_{_R}^{(L)} = a_{_R}^{(R)} = a_{_R} $ and $ a_{_T}^{(L)} = a_{_T}^{(R)} 
= a_{_T} $
\& accordingly as $ S\,=\,\left(\begin{array}{cc}
a_{_T} & a_{_R}\\a_{_R} & a_{_T}
\end{array}\right)$
it possesses the eigenvectors $\frac{1}{\sqrt{2}}\,\left(\begin{array}{c}
1\\1\end{array}\right)$ and $\frac{1}{\sqrt{2}}\,\left(\begin{array}{c}
1\\-1\end{array}\right)$ belonging to eigenvalues $ a_{_T} + a_{_R} $ 
\& $ a_{_T} - a_{_R} $ respectively \& hence with these eigenchannels as basis
the $ S $-matrix is diagonal with $ S\,=\,\left(\begin{array}{cc}
a_{_T} + a_{_R} & 0 \\ 0 & a_{_T} - a_{_R}
\end{array}\right)$. Noting that for $ V(x) = -\lambda \del(x) $ we have
non-trivial scattering only in the even channel. Thus $ a_{_T} - a_{_R} = 1 $
and $ a_{_T} + a_{_R} = A $ ( in the notation of the paper ). Therefore, 
$ a_{_T} = \frac{1}{2}\,( 1 + A )$. It may be noted that some authors 
\cite{datta} adopt a
different definition of the $S$-matrix from that we have taken.
 
\section{Location of the Resonance \& determination of the width}
It is best to look at $ A $ above $ E = \Del$ viz. both channels open \& 
track down its pole in $ k $ so that the sheet structure also becomes
clear. Eq.~(\ref{eq:wavampl}) gives for the pole
\begin{equation}\label{eq:poleA}
 q - i\frac{{\tilde \lam}}{2} + \frac{{\tilde \lam_c}^2}{4 k} = 0 \,\,. 
\end{equation}
At ${\tilde \lam_c} =0 $ the pole is at $ q = i\frac{{\tilde \lam}}{2} $  
or $ k^2 = {\tilde \Del} - \frac{{\tilde \lam}^2}{4} $ with 
${\tilde \Del} = \frac{2 m \Del}{\hbar^2} $ viz. $ k = \sqrt {
{\tilde \Del} - \frac{{\tilde \lam}^2}{4}}$.
Now switch on ${\tilde \lam_c}$ \& assuming ${\tilde \lam_c}$ is very small,
we may write
\begin{equation}\label{eq:pole}
 k = \sqrt {{\tilde \Del} - \frac{{\tilde \lam}^2}{4}} \, 
\bigl( 1 + \del_1 + i \del_2  \bigr)
\end{equation}
where $\del_1$ and $\del_2$ are small \& go smaller with smaller 
${\tilde \lam_c}$.\\
Thus,\,\,
$ q^2 = k^2 - {\tilde \Del} \simeq -\frac{{\tilde \lam}^2}{4} + \bigl(
{\tilde \Del} - \frac{{\tilde \lam}^2}{4}\bigr) \, 
\bigl(\del_1^2  - \del_2^2 + 2\del_1 + 2i\del_2 + 2i\del_1\del_2 \bigr)$ and
\begin{eqnarray}
q &=& i \,\frac{{\tilde \lam}}{2} \,\bigl[ 1 -  
\frac{2({\tilde \Del} - \frac{{\tilde \lam}^2}{4})}{{\tilde \lam}^2} \, \bigl(
\del_1^2  - \del_2^2 + 2\del_1 + 2i\del_2 + 2i\del_1\del_2 \bigr)\nn\\
&-&\frac{8 ({\tilde \Del} - \frac{{\tilde \lam}^2}{4})^2}{{\tilde \lam}^4} \, 
\bigl(\del_1^2  - \del_2^2 + 2i\del_1\del_2 \bigr) \bigr]  
\label{eq:poleA-qtru}
\end{eqnarray}
Here in taking the square-root of $ q^2 $ we have carefully taken the correct sign as with $ {\tilde \lam_c} \to 0 $ the pole must be at 
$ q = i\frac{{\tilde \lam}}{2}$.
\begin{eqnarray}
\frac{1}{k} &=& 
\frac{1}{\sqrt{{\tilde \Del} - \frac{{\tilde \lam}^2}{4}}}
\, \bigl( 1 + \del_1 +i \del_2 \bigr)^{-1} \nn\\
& \simeq & \frac{1}
{\sqrt{{\tilde \Del} - \frac{{\tilde \lam}^2}{4}}}\,
\bigl( 1 -\del_1 -i \del_2
+ \del_1^2 - \del_2^2 + 2 i \del_1 \del_2 + \cdots \bigr )
\label{eq:poleA-ktru}
\end{eqnarray}
Keeping terms up to the 2nd order in $\del$ in 
Expressions(~\ref{eq:poleA-qtru}) \& (~\ref{eq:poleA-ktru}) and then using 
them in Eq.(~\ref{eq:poleA})
we note that lowest terms in $\del$ ( with ${\tilde \lam_c}^2$ of the same 
order ) in imaginary \& real part of the Eq.(~\ref{eq:poleA}) yield
\begin{eqnarray}
-\frac{2({\tilde \Del} - \frac{{\tilde \lam}^2}{4})}{{\tilde \lam}}\,\del_1
-\frac{{\tilde \lam_c}^2}{4}\,\frac{1}{\sqrt{{\tilde \Del} - \frac{{\tilde \lam}^2}{4}}}
\,\del_2\,\,+ 8 \, \frac{({\tilde \Del} - \frac{{\tilde \lam}^2}{4})}
{{\tilde \lam}^4} \,{\tilde \Del}\,\frac{{\tilde \lam}}{2}\,\del_2^2 =0\nn\\
\frac{2({\tilde \Del} - \frac{{\tilde \lam}^2}{4})}{{\tilde \lam}}\,\del_2
+\frac{{\tilde \lam_c}^2}{4}\,\frac{1}{\sqrt{{\tilde \Del} - 
\frac{{\tilde \lam}^2}{4}}}\bigl( 1 - \del_1 \bigr) = 0 \,.\hspace{2cm}\nn
\end{eqnarray}
Note that while $\del_1$ is of $O({\tilde \lam_c}^4)$ , $\del_2$ is of 
$O({\tilde \lam_c}^2)$. From above two relations 
\begin{equation}\label{eq:imshift}
\del_2  = - {\tilde \lam_c}^2 \, \frac{{\tilde \lam}}
{8({\tilde \Del} - \frac{{\tilde \lam}^2}{4})^{3/2}}
\end{equation}
As $\del_1$ is of $O({\tilde \lam_c}^4)$ the imaginary part of the 
Eq.(~\ref{eq:poleA}) must be considered in a way so that it contains
$\del_1$ and $\del_2^2$ which are of ${\tilde \lam_c}^4$. Thus we get
\begin{equation}\label{eq:reshift}
\del_1  =  {\tilde \lam_c}^4 \,\, \frac{(2\,{\tilde \Del}\,+\,{\tilde \lam}^2)}
{64\,({\tilde \Del} - \frac{{\tilde \lam}^2}{4})^3}
\end{equation}
Using expressions (~\ref{eq:reshift}) \& (~\ref{eq:imshift}) we find the
pole position from Eq.(~\ref{eq:pole}) as
\begin{eqnarray}\label{eq:pole-k}
k_p \cong \sqrt{{\tilde \Del} - \frac{{\tilde \lam}^2}{4}} \,
+ {\tilde \lam_c}^4 \, \frac{(2\,{\tilde \Del}\,+\,{\tilde \lam}^2)}
{64\,({\tilde \Del} - \frac{{\tilde \lam}^2}{4})^{5/2}}
-i \, {\tilde \lam_c}^2 \, \frac{{\tilde \lam}}
{8({\tilde \Del} - \frac{{\tilde \lam}^2}{4})}\\
\nn\\
E_p -i\,\frac{\Ga}{2} \cong \frac{\hbar^2 k_p^2}{2 m } \simeq \Del - 
\frac{ \hbar^2 }{2 m }\,\left[\frac{{\tilde \lam}^2}{4} \, -
 {\tilde \lam_c}^4 \, \frac{({\tilde \Del} + \frac{{\tilde \lam}^2}{4})}
{16({\tilde \Del} - \frac{{\tilde \lam}^2}{4})^2}\right] 
-i \, \frac{ \hbar^2 }{2 m }\,{\tilde \lam_c}^2 \, 
\frac{{\tilde \lam}}{4({\tilde \Del} - \frac{{\tilde \lam}^2}{4})^{1/2}}
\label{eq:pole-E}
\end{eqnarray}  
which are referred as Eq.(~\ref{eq:pole-del-k}) \& Eq.(~\ref{eq:pole-del-E})
in the text.

\end{document}